\documentstyle[prd,aps,preprint,epsfig]{revtex}

\newcommand{\acoup}{ {a_{\psi_{z} \tilde{\phi} \phi}} }

\newcommand{\gev}{ {\rm GeV} }

\newcommand{\kakko}[1]{{\left({#1}\right)}}

\newcommand{\mgrav}{ {m_{3/2}} }

\newcommand{\mev}{ {\rm MeV} }
\newcommand{\mG}{ {M_{\rm G}} }
\newcommand{\mphi}{ {m_{\phi}} }
\newcommand{\mmodulino}{ {m_{\tilde{\phi}}} }
\newcommand{\order}{{\cal O}}

\newcommand{\tev}{ {\rm TeV} }
\newcommand{\tenten}{ {\cdot\cdot\cdot} }
\newcommand{\g}{3/2}
\newcommand{\bea}{\begin{eqnarray}}  \newcommand{\eea}{\end{eqnarray}}
\newcommand{\beq}{\begin{equation}}  \newcommand{\eeq}{\end{equation}}
\newcommand{\bef}{\begin{figure}}  \newcommand{\eef}{\end{figure}}
\newcommand{\bec}{\begin{center}}  \newcommand{\eec}{\end{center}}
  
\newcommand{\lmk}{\left(}  \newcommand{\rmk}{\right)}
\newcommand{\lkk}{\left[}  \newcommand{\rkk}{\right]}

\newcommand{\infl}{{\rm inf}}
\newcommand{\th}{{\rm th}}
\newcommand{\mphicr}{m_{\phi \rm cr}}
\newcommand{\Jcr}{J_{\rm cr}}

\def\PLB#1#2#3{Phys. Lett. B {\bf #1}, #2 (19#3)}

\def\PRDD#1#2#3{Phys. Rev. D {\bf #1}, #2 (20#3)}

\begin{document}
\tighten

\preprint{
\noindent
\begin{minipage}[t]{3in}
\begin{flushright}
OU-TAP-227\\
TU-711 \\
hep-ph/0403043 \\
\end{flushright}
\end{minipage}
}

\title{Production and dilution of gravitinos by modulus decay}
\author{Kazunori~Kohri$^1$, Masahiro~Yamaguchi$^2$, and Jun'ichi~
Yokoyama$^1$}

\address{$^1$Department of Earth and Space Science, Graduate School of
Science,\\ Osaka University,
Toyonaka  560-0043, Japan}

\address{$^2$Department of Physics, Tohoku University, Sendai
980-8578, Japan}

\maketitle
\begin{abstract}
    We study the cosmological consequences of generic scalar fields
    like moduli which decay only through gravitationally suppressed
    interactions.  We consider a new production mechanism of
    gravitinos from moduli decay, which might be more effective than
    previously known mechanisms, and calculate the final
    gravitino-to-entropy ratio to compare with the constraints imposed
    by successful big bang nucleosynthesis (BBN) etc., taking possible
    hadronic decays of gravitinos into account. We find the modulus
    mass smaller than $\sim 10^4$ TeV is excluded. On the other hand,
    inflation models with high reheating temperatures $T_{R,\rm inf}
    \sim 10^{16}$ GeV can be compatible with BBN thanks to the
    late-time entropy production from the moduli decay if model
    parameters are appropriately chosen.
\end{abstract}



\section{Introduction}
\label{sec:introduction}

One of the consequences of local supersymmetry or supergravity is the
existence of gravitinos, the superpartner of gravitons, whose natural
mass scale is the weak scale $\order (0.1-10)$ TeV.  If they are not
the lightest supersymmetric particle (LSP), they decay into LSP and
other  high energy particles such as photons, neutrinos,
quark-antiquark pairs or gluons  after/during the  big-bang
nucleosynthesis (BBN) epoch. Such decay products may change the light
element abundances by  changing the baryon-to-entropy,
neutron-to-proton ratios, or destroying the produced elements,  which
may result in a significant discrepancy between yields and
observations. This is called  the ``gravitino problem''. A lot of
authors have studied a variety of constraints on the decaying
gravitinos from
BBN~\cite{Weinberg:zq,Krauss:1983ik,Lindley:1984bg,Khlopov:pf,Ellis:1984eq,Juszkiewicz:gg,Ellis:1984er,Audouze:be,Kawasaki:1986my,Scherrer:1987rr,Dominguez:1987,Reno:1987qw,Dimopoulos:1987fz,Ellis:1990nb,Kawasaki:1994af,Kawasaki:1994bs,Protheroe:dt,Holtmann:1998gd,Jedamzik:1999di,Kawasaki:2000qr,Kohri:2001jx,Cyburt:2002uv,Kawasaki:2004yh}~\footnote{
Cosmological constraints on stable gravitinos are studied in
Refs.~\cite{Pagels:ke,Berezinsky:kf,Moroi:1993mb} and references
therein.
}. 

In the standard inflationary cosmology, gravitinos are produced by the
scattering processes of thermal particles at the reheating epoch just
after inflation.  The yield parameter $y_{\g}$ of gravitinos, which is
the  ratio of  their number density to entropy density,
$y_{\g}=n_{\g}/s$, is approximately given as a function of the
reheating temperature after inflation, $T_{R,\rm inf}$, as
\cite{Kawasaki:1994af,Moroi:1995fs,Bolz:2000fu}~\footnote{
The error of this approximation formula is  within  $\sim 5\%$ for
$T_{R} = 10^{6}$ -- $10^{14}$ GeV, and $\sim 25\%$ for $T_{R} =
10^{2}$ -- $10^{19}$ GeV~\cite{Moroi:1995fs}.
}
\begin{eqnarray}
    \label{eq:y32_scattinf}
    y_{3/2,\rm inf} \simeq 1.5 \times 10^{-12} 
             \left(\frac{T_{R,\rm inf}}{10^{10} \gev}\right) .
\end{eqnarray}
Hence it is customary to express the constraint on their abundance
imposed by BBN as that on the reheating temperature $T_{R,\rm inf}$.
For example, if gravitino mass is equal to 0.1 TeV  we find an upper
bound   $T_{R,\rm inf}\sim 10^6$GeV,  which imposes a constraint on
model building of inflation.  For example, hybrid inflation
\cite{hybrid} is difficult to reconcile with this low reheating
temperature, in which the inflaton is typically coupled to gauge
fields and decays rapidly.  So it is preferable to have another
inflation after hybrid inflation as suggested by recent observational
data \cite{YY}.

We would like to point out, however, that these constraints on 
$T_{R,\infl}$ 
have all been obtained under the implicit assumption that $y_{\g,
\infl}$ remains constant until their lifetime,
\begin{eqnarray}
    \label{eq:lifetime_grav}
    \tau_{3/2} \simeq 4 \times 10^{5} \sec \frac1{N_{\rm G}}
    \left(\frac{\mgrav}{1 \ 
    \tev}\right)^{-3},
\end{eqnarray}
where we assume that  the gravitino decays into a massless gauge boson
and a gaugino, $N_{\rm G}$ is the number of the generators of the
gauge group, and $m_{3/2}$ is  the gravitino mass.  In this paper we
argue that both the denominator and the numerator of $y_{\g}=n_{\g}/s$
are subject to change between the reheating epoch after inflation and
their decay time, apart from the dilution due to cosmic expansion
which does not change the ratio $y_{\g}$ itself.  We then derive more
appropriate constraints imposed by BBN.

In supergravity or superstring theories there appear a number of
long-lived scalar fields which decay only through gravitational
interactions, such as moduli, dilatons, or Polonyi field, which are
referred to collectively as the modulus $\phi$ hereafter.  The modulus
$\phi$ starts coherent field oscillation to dissipate its energy
density as the Hubble parameter $H$ becomes smaller than its mass
$\mphi$. Because its dissipation rate is smaller than that of
radiation, the universe turns to be matter dominated well before the
lifetime of $\phi$, when it decays producing not only huge amount of
entropy, causing what is called the moduli problem
\cite{Coughlan:1983ci,Banks:1993en,deCarlos:1993jw},  but also
undesirable particles for cosmology.   So far a number of groups have
studied cosmological constraints on the modulus decays depending on
the properties of each decay product, {\it e.g.}, for LSPs not to
close the universe~\cite{MYY,Kawasaki:1995cy,Moroi:1999zb}, and  for
radiations  to complete thermalization~\cite{Kawasaki:1999na}.

The effects of decaying moduli on the gravitino problem are twofold.
One is that the entropy produced by their decay dilutes primordial
gravitino abundance, which is a good news to relax constraints on
inflation model building.  The other is that these unwanted particles
may also be  produced directly by the decay of moduli $\phi$.
Including these two effects, it is not apparent for us to find the
allowed region of model parameters to avoid both the gravitino problem
and moduli problem. So far no one has considered this type of scenario
with decaying moduli. Therefore, we comprehensively study the effects
in this work.

The constraint by gravitinos produced directly by the decay of moduli
was first studied by Hashimoto et  al.\ \cite{Hashimoto:1998mu}, who
considered the case that the  modulus decays into two gravitinos. In
the present paper, however, we point out a more efficient mechanism of
direct gravitino production from moduli decay, which is the mode that
$\phi$ decays into its superpartner modulino, $\tilde{\phi}$, and a
gravitino, $\psi_{\g}$, namely, $\phi \to \tilde{\phi} + \psi_{3/2}$.  
We
incorporate this decay mode and consider cosmological constraints on
the masses of gravitinos and moduli from BBN and other observations.

The fate of the decaying moduli depends on  its mass $m_{\phi}$. In
\cite{Banks:1993en,deCarlos:1993jw}, it was presumed that moduli
fields acquire masses through supersymmetry breaking and then their
masses are comparable to masses of superparticles. In this case the
modulus field is long-lived and the reheating at the modulus decay
takes place with the temperature much below 1MeV, spoiling the success
of the BBN. (See, however, \cite{MYY}).  Recently it was realized that
a mechanism to stabilize the moduli fields is operative in the
compactification with non-zero NS and RR fluxes in  certain string
theories~\cite{Giddings}. This makes most of the moduli fields very  massive,
typically around the string scale.  Still there are  some moduli which
are not stabilized. In particular, the modulus field which is
responsible to determine the size of the compactification is not
stabilized in the flux compactification. With the ignorance of
possible mechanisms on mass generation, we take the mass of the
modulus field as a free parameter in this paper.

\section{Modulus decay into gravitinos}

The relevant terms in the supergravity Lagrangian which describe the
decay mode of our concern are given by
\begin{eqnarray}
    \label{eq:Lag_decay}
    {\cal L}  &=& \left[\acoup\frac{1}{\mG} Z \Phi^{\dagger}
    \Phi\right]_{\theta \theta \bar{\theta} \bar{\theta}}.
\end{eqnarray}
This is allowed by gauge invariance. Furthermore it has the same
structure as a source of soft supersymmetry (SUSY) breaking terms of squarks
and sleptons in gravity mediation of SUSY breaking (see {\em
  e.g.} \cite{Kaplunovsky:1993rd}).
Here $\acoup$ is the dimensionless coupling constant,  $Z = z +
\sqrt{2}\theta \psi_{z} + \theta \theta F_{z} + \tenten$ is the chiral
superfield which breaks SUSY by acquiring an $F$ term. Therefore, it
includes goldstino $\psi_{z}$ or the longitudinal mode of gravitino,
with a complex scalar field $z$ and an auxiliary field $F_{z}$ while
$\theta$ and $\bar{\theta}$ are  two-component Grassmann
variable. $\Phi=\phi + \sqrt{2}\theta
\tilde{\phi} + \theta \theta F_{\phi} +i \sqrt{2} \theta
\partial_{\mu} \tilde{\phi} \theta \sigma^{\mu} \bar{\theta}
+\tenten$ is the chiral superfield 
 which includes the modulus $\phi$, and the modulino
$\tilde{\phi}$  with an
auxiliary field $F_{\phi}$. Then we obtain

\begin{eqnarray}
    \label{eq:Lag_decay2}
    {\cal L} &=& \acoup \frac1{\mG} F_{z} F_{\phi}^{\dagger} \phi 
    + \acoup \frac1{\mG} \psi_{z} \left(i \sigma^{\mu} \partial_{\mu}
    \overline{\tilde{\phi}} \right)  \phi \nonumber + \tenten \\
    &=& -\sqrt{3}\acoup \ \mgrav \mmodulino \phi \phi 
    + \acoup \frac{\mmodulino}{\mG} \psi_{z} \tilde{\phi}  \phi +
    \tenten, 
\end{eqnarray}
where we have incorporated the SUSY breaking effect, $\langle F_{z}
\rangle = \sqrt{3} \mgrav \mG$ with 
the reduced Planck mass $\mG$ ($\cong 2.44 \times 10^{18}\ \gev$)
without generating a cosmological constant. We also used the field
equations of motion, for the modulino $i \sigma^{\mu} \partial_{\mu}
\overline{\tilde{\phi}} - \mmodulino \tilde{\phi} = 0$ with its mass
$\mmodulino$, and for the modulus $F_{\phi}^{\dagger} + \mmodulino
\phi= 0$.

Then, the mass matrix of $\phi$  is given by
\begin{eqnarray}
    \label{eq:mass-matrix}
    {\cal M} = \left(
      \begin{array}{cc}
          \mmodulino^{2} & \sqrt{3} \acoup \mgrav \mmodulino \\
          \sqrt{3} \acoup\mgrav \mmodulino & \mmodulino^{2} 
      \end{array}
    \right).
\end{eqnarray}
Therefore, we can diagonalize it and get two eigen values
$\mphi_{\pm}$, which are
\begin{eqnarray}
    \label{eq:eigen_values}
    \mphi_{\pm} = \sqrt{\mmodulino^{2} \pm \sqrt{3} \acoup \ \mgrav
    \mmodulino}. 
\end{eqnarray}
We find that the decay $\phi \to \tilde{\phi} + \psi_{3/2}$ is
kinematically allowed if $\acoup >
\kakko{2+\mgrav/\mmodulino}/\sqrt{3}$. In this paper, we consider the
case $\mphi_{+} - \mphi_{-} \ll \mmodulino$ or $\mmodulino \gg
\sqrt{3}\acoup \ \mgrav$, so that the field oscillation of $\phi_{+}$
continues as long as that of $\phi_{-}$. Hereafter for simplicity we
assume that only $\phi_{+}$ is present and induces a coherent
oscillation with mass,
\begin{eqnarray}
    \label{eq:eigen_values2}
    \mphi \simeq \mmodulino + \frac{\sqrt{3}}{2} \acoup \ \mgrav.
\end{eqnarray}


As is seen in (\ref{eq:Lag_decay2}), the relevant term of the Lagrangian
that describes the decay reads
\begin{eqnarray}
    \label{eq:Lint}
    {\cal L}_{\rm int} = \acoup \  \frac {m_{\tilde{\phi}}}{\mG}
 \psi_{z} \tilde{\phi} \phi \equiv h \psi_{z} \tilde{\phi} \phi.
\end{eqnarray}
Then the decay rate of the modulus $\phi$ into  the gravitino
$\psi_{z}$ and the modulino $\tilde{\phi}$ is given by
\begin{eqnarray}
    \label{eq:rate_modulino}
    \Gamma(\phi \to \tilde{\phi} + \psi_{z}) = \frac{h^2}{8 \pi}m_{\phi}
    \left[1-\frac{\left(\mmodulino+\mgrav\right)^{2}}{\mphi^{2}} \right]
^{3/2}
    \left[1-\frac{\left(\mmodulino-\mgrav\right)^{2}}{\mphi^{2}} \right]
^{1/2}.
\end{eqnarray}
Because we are assuming  $\mphi \gg \sqrt{3}\acoup \ \mgrav$, we obtain
the following approximate formula for Eq.~(\ref{eq:rate_modulino}),
\begin{eqnarray}
    \label{eq:rate_modulino2}
    \Gamma(\phi \to \tilde{\phi} + \psi_{z}) 
    &\simeq& \frac{A_{4}}{2\pi}\frac{\mphi \mgrav^{2}}{\mG^{2}},
\end{eqnarray}
where the factor $A_{4}$ is given by
\begin{eqnarray}
    A_{4} &\equiv& \acoup^{2}\left(\frac{\sqrt{3}}{2}\acoup-1\right)
    \sqrt{\frac32\acoup^{2}-1}  \\
    &\simeq& \ \frac{3\sqrt{2}}{4}\acoup^{4}
    \qquad \qquad {\rm for}\   \acoup \gg 1.
\end{eqnarray}
Note that this decay rate is very sensitive to the coupling constant
$\acoup$. Therefore, for relatively larger $\acoup$, the decay width
of the mode $\phi \to \psi_{3/2} + \tilde{\phi}$ can become larger
than that of the mode $\phi \to 2 \psi_{3/2}$ which was studied in
Ref.~\cite{Hashimoto:1998mu}~\footnote{
The decay width of the mode $\phi \to 2 \psi_{3/2}$ is proportional to
the square of the relevant coupling constant as opposed to $\propto
\acoup^{4}$ here. On the other hand, if the coupling constant is
smaller than unity or the decay mode $\phi \to \psi_{3/2} +
\tilde{\phi}$ is kinematically forbidden, the decay mode $\phi \to 2
\psi_{3/2}$ is more effective.
}. 

On the other hand, the decay width of modulus
into radiation is represented by
\begin{eqnarray}
    \label{eq:total}
    \Gamma(\phi \to {\rm radiations}) = N \frac{\mphi^{3}}{\mG^{2}},
\end{eqnarray}
where $N$ depends on the number of the final states. For example, if
all of the particle contents in the minimal supersymmetric standard
model (MSSM) appear in the final states, we approximately obtain $N =
\order(10)$~\cite{Moroi:1999zb}. Because we are interested in the
parameter space which satisfies $\mphi \gg \sqrt{3}\acoup \mgrav$, the
above decay width into radiation is the dominant decay mode, which we
identify with the total width, $\Gamma_{\rm tot}$ hereafter.
Thus the branching ratio to gravitino production reads
\begin{eqnarray}
    \label{eq:B32}
    B_{3/2} \equiv
    \Gamma(\phi \to \tilde{\phi} + \psi_{z})/\Gamma_{\rm tot} 
    = \frac{A_{4}}{2\pi N}\left(\frac{\mgrav}{\mphi}\right)^{2}.
\end{eqnarray}

\section{Gravitino abundance after modulus decay}
\label{sec:gra-prod}

Now we consider cosmological evolution of the modulus $\phi$.  By
adding an appropriate constant we redefine $\phi$ so that it has a
global minimum at $\phi=0$ and assume that the mass term dominates its
potential energy density for simplicity.  We also assume that its
initial amplitude, $\phi_i$, is of order of $\mG$ or smaller.  The
modulus remains there until the Hubble parameter $H$ decreases to
$m_\phi$, when it starts coherent oscillation around the origin.  As
the field oscillation red-shifts less rapidly than radiation, the
universe will be dominated by $\phi$ at the time, \beq t_{eq}\cong
\frac{9}{2m_\phi}\lmk\frac{\mG}{\phi_i}\rmk^4.  \eeq After that time,
the  expansion law of the cosmic scale factor $a(t)$ is the same as
that in matter dominated regime, $a(t)\propto t^{2/3}$, until $\phi$
decays at $t_d\simeq \Gamma_{\rm tot}^{-1} $, when the ratio of energy
density of $\phi$ to that of radiation reads \beq \frac{\rho_\phi
(t_d)}{\rho_R(t_d)}=\lmk\frac{t_d}{t_{eq}}\rmk^{2/3} \cong
\lkk\frac{2}{9N}\lmk\frac{\mG}{m_\phi}\rmk^2
\lmk\frac{\phi_i}{\mG}\rmk^4\rkk^{2/3}.  \eeq Thus the entropy
increase factor, $\Delta$, reads
\begin{eqnarray}
    \label{eq:dilution}
    \Delta \equiv \frac{s_{\rm after}(t_{d})}{s_{\rm
           before} (t_{d})} 
           = \kakko{\frac{\rho_{\phi}(t_{d})}{\rho_{R} (
           t_{d})}}^{3/4} 
           = \sqrt{\frac{2}{9N}}
           \frac{\mG}{\mphi}\kakko{\frac{\phi_{i}}{\mG}}^{2},
\end{eqnarray}
where $s_{\rm after}(t_{d})$ ( $s_{\rm before} (t_{d})$ ) is the
entropy density just after (before) the modulus decay. In this paper
we only consider the case that $\Delta \gg 1$.

Using an approximation that the modulus energy density is fully
converted to radiation when $\Gamma_{\rm tot} = H(T_{R})$, 
the reheating temperature after the modulus decay is found to be
\begin{eqnarray}
    \label{eq:TR}
    T_{R} &=& \left(\frac{90}{\pi^{2}g_{*}}\right)^{1/4} 
    \sqrt{\Gamma_{\rm tot} \mG} 
  =\lmk\frac{90N^2}{\pi^2g_*}\rmk^{1/4}\mphi^{3/2} \mG^{-1/2}
\nonumber \\
    &=& 11 N^{1/2}\lmk\frac{g_*}{10^2}\rmk^{-1/4}
\lmk\frac{m_\phi}{10^6 \rm TeV}\rmk^{3/2} ~{\rm TeV}.
\end{eqnarray}
Here $g_{*}$ denotes total effective numbers 
of relativistic degrees of freedom.  We find $g_*=10.75$ just before the
onset of BBN and $g_*=228.75$ if all the particle contents of the 
minimal
supersymmetric standard model are massless and in thermal equilibrium.
We note that the reheating temperature should satisfy
\begin{eqnarray}
    \label{eq:lower_bound}
    T_{R} > 1.2 \ \mev, \qquad ({\rm at} \ 95 \% \ {\rm C.L.}),
\end{eqnarray}
so that the  neutrino background can complete thermalization to warrant
successful BBN~\cite{Kawasaki:1999na}~\footnote{
In Ref.~\cite{Kawasaki:1999na}, however,  $T_{R}$ is defined by
$\Gamma_{\rm tot} =3 H(T_{R})$. Using their definition, the lower
bound of $T_{R}$ turns into 0.7 MeV, which they reported. On the other
hand, note that they also  discussed constraints for emitted hadrons
by decaying modulus not to influence on the neutron to proton ratio
before/during the BBN epoch. Then, the lower bound is pushed to the
severer one ($T_{R} > 5.2$ MeV). In this paper, however,  we do not go
into such specifics. Here we adopt the conservative one ($T_{R} > 1.2$
MeV).
}.
This means that $m_\phi$ should satisfy\footnote{
This constraint may be evaded if the Universe underwent late-time
inflation such as the thermal inflation to dilute the energy
density of the modulus field \cite{LythStewart}. 
In this case, the reheating temperature after the thermal inflation
should satisfy this limit.
}
\beq
  m_\phi > 11N^{-1/3}\lmk\frac{g_*}{10.75}\rmk^{1/6} {\rm TeV}.
\eeq

Now there are three sources of gravitinos after the modulus decay.
One is the primordial gravitinos which were produced just after
inflation and diluted by the entropy from moduli.  Using
(\ref{eq:y32_scattinf}),  its abundance is given by \beq
\label{inflation}
  y_{3/2,\infl}(t_d)=\frac{y_{3/2,\infl}(t_i)}{\Delta} =
1.4\times 10^{-21}N^{1/2}
\lmk\frac{\mphi}{10^6 \rm TeV}\rmk
\lmk\frac{\phi_i}{\mG}\rmk^{-2}
\left(\frac{T_{R,\rm inf}}{10^{10} \gev}\right),
\eeq
where $y_{3/2,\infl}(t_i)$ refers to the primordial value after
inflation and we have
assumed no significant entropy production took place between reheating
after inflation and modulus decay.

On the other hand, the yield parameter due to direct production of
gravitinos from $\phi$ is given by
\begin{eqnarray}
    \label{direct}
    y_{3/2,\phi}(t_d) &\equiv& B_{3/2}\frac{\rho_{\phi}/\mphi}{s}
           = \frac{3A_{4}}{8\pi N}\lmk\frac{\mgrav}{\mphi}\rmk^2
\frac{T_R}{\mphi} \nonumber \\
&=& 1.1\times 10^{-16}\lmk\frac{\acoup}{3}\rmk^4N^{-1/2}
\lmk\frac{g_*}{10^2}\rmk^{-1/4}\lmk\frac{\mgrav}{1 \rm TeV}\rmk^2
\lmk\frac{\mphi}{10^6 \rm TeV}\rmk^{-3/2}.
\end{eqnarray}
Here we have used the following approximate relations; $s =
(2\pi^{2}/45)g_{*}T_{R}^{3}$, and
$\rho_{\phi}\simeq\rho_{R}=(\pi^{2}/30)g_{*}T_{R}^{4}$ with the energy
density of radiation $\rho_{R}$ at the reheating time.

Finally, gravitinos are also produced by the scattering
process in the thermal bath at the reheating due to modulus decay, 
whose contribution to the yield parameter reads
\begin{eqnarray}
    \label{thermal}
    y_{3/2,\rm th}(t_d) &\simeq& 1.5 \times 10^{-12} 
             \left(\frac{T_{R}}{10^{10} \gev}\right) \nonumber \\
     &=& 1.8 \times 10^{-18}N^{1/2}
\lmk\frac{g_*}{10^2}\rmk^{-1/4}\lmk\frac{\mphi}{10^6 \rm TeV}\rmk^{3/2}.
\end{eqnarray}

Comparing these three equations with each other we can find which
production mechanism is dominant for each combination of model
parameters.
We first calculate which of the late time production mechanism is more
efficient. From (\ref{direct}) and (\ref{thermal}) we find the direct
production from decaying moduli is more efficient than the thermal
scattering processes if
\beq
 \mphi < 4\times 10^6 \lmk\frac{\acoup}{3}\rmk^{4/3}N^{-1/3}
\lmk\frac{\mgrav}{1 \rm TeV}\rmk^{2/3} {\rm TeV}\equiv \mphicr.
\eeq
Then comparing $y_{3/2,\rm th}(t_d)$ with $y_{3/2,\infl}(t_d)$ for the
case $\mphi > \mphicr$ and  $y_{3/2,\phi}(t_d)$ 
with $y_{3/2,\infl}(t_d)$ for the opposite
case $\mphi < \mphicr$, we find that primordial gravitinos can dominate
over the late-time counterparts only if the inequality 
\beq
\label{J}
  J\equiv \lmk\frac{\phi_i}{\mG}\rmk^{-2}
\left(\frac{T_{R,\rm inf}}{10^{10} \gev}\right)
> 3\times 10^3  \lmk\frac{\acoup}{3}\rmk^{2/3}N^{-1/6}
\lmk\frac{g_*}{10^2}\rmk^{-1/4}\lmk\frac{\mgrav}{1 \rm
TeV}\rmk^{1/3}\equiv \Jcr,
\eeq
is satisfied, where the factor $J$ is determined by the combination of
efficiency of primordial production of gravitinos
and deficiency of dilution due to entropy production from $\phi$.

First suppose that the inequality (\ref{J}) is satisfied.
Then we find that the direct production from
 $\phi$ is dominant for
\beq
\mphi <2\times 10^6 \lmk\frac{\acoup}{3}\rmk^{8/5}N^{-2/5}
\lmk\frac{g_*}{10^2}\rmk^{-1/10}\lmk\frac{\mgrav}{1 \rm TeV}\rmk^{4/5}
\lmk\frac{J}{10^4}\rmk^{-2/5} {\rm TeV},
\eeq
while primordial one contributes the most for
\bea
&&2\times 10^6 \lmk\frac{\acoup}{3}\rmk^{8/5}N^{-2/5}
\lmk\frac{g_*}{10^2}\rmk^{-1/10}\lmk\frac{\mgrav}{1 \rm TeV}\rmk^{4/5}
\lmk\frac{J}{10^4}\rmk^{-2/5}  {\rm TeV}\nonumber \\
&&~~~~~~~~~~~~~~~~~~~~~ < \mphi < 6\times
10^7\lmk\frac{g_*}{10^2}\rmk^{1/2} \lmk\frac{J}{10^4}\rmk^{2}  {\rm
TeV}.
\eea
For the case
\beq%
6\times 10^7\lmk\frac{g_*}{10^2}\rmk^{1/2}
\lmk\frac{J}{10^4}\rmk^{2}  {\rm TeV} < \mphi,
\eeq%
the thermal scattering in the plasma produced by modulus decay is the
most important.

The situation with $J < \Jcr$ is much simpler.  We find that
$y_{3/2,\phi}(t_d)$  is dominant if $\mphi <\mphicr$  and that
$y_{3/2,\th}(t_d)$ is the largest for $\mphi >\mphicr$.

\section{Cosmological constraints on decaying gravitinos}
\label{sec:photodiss}

Having fully analyzed the yield parameter for all the possible cases
in our model, we are now in a position to impose cosmological
constraints on our model parameters.  First we summarize the
constraints we use.

If the gravitino further decays into lighter particles such as
photons, neutrinos, gluons, or quark-antiquark pairs, they induce
electromagnetic or hadronic showers which would influence on BBN
because they may change the light element abundances by destroying
them or changing the  neutron to proton ratio. So far a number of
authors have studied a  variety of constraints on decaying gravitinos
from
observations~\cite{Weinberg:zq,Krauss:1983ik,Lindley:1984bg,Khlopov:pf,Ellis:1984eq,Juszkiewicz:gg,Ellis:1984er,Audouze:be,Kawasaki:1986my,Scherrer:1987rr,Dominguez:1987,Reno:1987qw,Dimopoulos:1987fz,Ellis:1990nb,Kawasaki:1994af,Kawasaki:1994bs,Protheroe:dt,Holtmann:1998gd,Jedamzik:1999di,Kawasaki:2000qr,Kohri:2001jx,Cyburt:2002uv,Kawasaki:2004yh}.
Here  we consider the following  two typical examples. One is the case
that the gravitino decays only into a photon and an LSP neutralino
$\chi$ with $N_{G} = 1$. Even in this case, there also exists a
hadronic decay mode into quark-antiquark pairs with the hadronic
branching ratio $B_{H} \sim \alpha/(4 \pi) \sim 10^{-3}$. The other is
that the gravitino decays only into a gluon $g$ and a gluino
$\tilde{g}$ with $N_{G} = 8$. Then, the hadronic branching ratio is
unity $B_{H}= 1$.  In both cases, the electromagnetic branching ratio
is practically unity because almost all energies of their decay
products are immediately turned into high energy radiations that
participate in the photodissociation of light elements. Including more
hadronic decay modes in addition to the radiative decay mode, it is
known that the resultant constraints get more
stringent~\cite{Dominguez:1987,Reno:1987qw,Dimopoulos:1987fz,Kohri:2001jx,Kawasaki:2004yh}.

Compared with observational light element abundances, we get the upper
bound on $y_{3/2}$ as a function of the gravitino mass. It is
approximately expressed by~\cite{Kohri:2001jx,Kawasaki:2004yh}
\begin{eqnarray}
    \label{eq:y32_upper_photodis}
    y_{3/2} \lesssim \left\{ 
      \begin{array}{ll}
          3 \times 10^{-16} - 6 \times 10^{-15}
          \ (3 \times 10^{-18} - 2 \times 10^{-16})
          \ &{\rm for} \  \mgrav = 0.1 - 1 \tev, 
          \\
          2 \times 10^{-15} - 2 \times 10^{-13}
          \ (3 \times 10^{-18} - 2 \times 10^{-13})
          \ & {\rm for} \  \mgrav = 1 \tev - 10 \tev,
          \\
          2 \times 10^{-13} - \sim 10^{-9}
          \ (1 \times 10^{-13} - 2 \times 10^{-13})
          \ & {\rm for} \  \mgrav = 10 \tev - 30 \tev,
      \end{array}
    \right.
\end{eqnarray}
for $B_{H} = 10^{-3}$ ($B_{H} = 1$).  Of course, if we do not consider
the decaying moduli, (\ref{eq:y32_upper_photodis}) is transformed into
upper bounds on the reheating temperature after the primordial
inflation $T_{R,\rm inf}$ as follows.
\begin{eqnarray}
    \label{eq:y32_upper_tr_inf}
    T_{R,\rm inf} \lesssim \left\{ 
      \begin{array}{ll}
           2 \times 10^{6} - 4 \times 10^{7}
          \quad (2 \times 10^{4} - 1 \times 10^{6})
          \quad &{\rm for} \  \mgrav = 0.1 - 1 \tev, 
          \\
          1 \times 10^{7} - 1 \times 10^{9}
          \quad (2 \times 10^{4} - 2 \times 10^{9})
          \quad & {\rm for} \  \mgrav = 1 \tev - 10 \tev,
          \\
          1 \times 10^{9} - \sim 10^{13}
          \quad (1 \times 10^{9} - 2 \times 10^{9})
          \quad & {\rm for} \  \mgrav = 10 \tev - 30 \tev,
      \end{array}
    \right.
\end{eqnarray}
for $B_{H} = 10^{-3}$ ($B_{H} = 1$).  Satisfying these upper bounds is
just one solution to avoid the gravitino problem without any late-time
entropy production by  decaying particles such as moduli. For more
detail, see the  results in
Ref.~\cite{Kohri:2001jx,Kawasaki:2004yh}~\footnote{
The radiative decay of gravitinos also influence on the shape of
Planck distribution of cosmic microwave background  (CMB) radiation
through the $\mu$ or $y$ distortion.  However, this type of limit is
weaker than that of the photodissociation
process~\cite{Ellis:1990nb,Holtmann:1998gd}.
}. 

In the present scenario, an LSP is produced by each gravitino decay. 
Therefore we should
consider the constraints on relic density of LSPs. We can relate their
number density  $n_{\rm LSP}$ with that of gravitinos,
\begin{eqnarray}
    \label{eq:yLSP}
    y_{3/2} =  \frac{n_{\rm LSP}}{s} = \frac{\Omega_{\rm LSP}}{m_{\rm
    LSP}}    \frac{\rho_{\rm cr}}{s}, 
\end{eqnarray}
where $m_{\rm LSP}$ and $\Omega_{\rm LSP}$ are the mass and  the
density parameter of LSPs, respectively,  and $\rho_{\rm cr}$ is the
critical density in the universe.\footnote{ In our scenario, the
modulus decay also produces a modulino, followed by the decay into a
LSP. Thus the total amount of the LSPs could be up to twice as large
as the right-hand-side of Eq.~(\ref{eq:yLSP}). We do not include this
effect in our analysis, which does not affect our conclusions.}  From
observation of CMB anisotropies, the WMAP collaboration reported that
$\Omega_{\rm LSP} < 0.35 $ at 95 \% C.L.~\cite{Spergel:2003cb}. Here,
$\rho_{\rm cr}/s$ is equal to the present value $\rho_{{\rm
cr}\hspace{0.15pc}0}/s_{0}$, where $\rho_{\rm cr \hspace{0.15pc} 0} =
4.2\times 10^{-47} \gev^{4}(h/0.72)^{2}$ with $h = 0.72 \pm
0.05$~\cite{Spergel:2003cb,freedman:2001}, and $s_{0} = 2.2\times
10^{-38} \gev^{3} (T_{0}/2.725 {\rm K})^{3}$ with the present value of
the photon temperature $T_{0} = 2.725 \pm 0.002 {\rm
K}$~\cite{Mather:1999}. Normalizing at $m_{\rm LSP} = 100 \ \gev$, we
get the upper bound on $y_{3/2}$ from Eq.~(\ref{eq:yLSP}),
\begin{eqnarray}
    \label{eq:yLSP_upper}
    y_{3/2} < 6.6 \times 10^{-12} 
    \left(\frac{m_{\rm LSP}}{100 \ \gev}\right)^{-1} \equiv
     y_{3/2}^{\rm max},
\end{eqnarray}
which is a conservative bound that turns out to be exact in the case
thermal relic density is negligibly small.\footnote{In fact, as the
modulus mass increases, the reheating temperature exceeds the
freeze-out temperature  above  which the thermal
production/annihilation of the LSPs is effective. In  this  case a
significant contribution to $\Omega_{\rm LSP}$ may come from the
thermal production soon after the moduli decay. This contribution is
very  sensitive to the SUSY mass spectrum and, for simplicity, we do
not include it in  this paper.}  Here we have assumed that the
abundance of these neutralinos produced by such decaying gravitinos
does not change though possible annihilation process, which can be
shown to be ineffective as follows. The annihilation cross section of
the neutralino $\sigma_{\rm ann}$ is a complicated function of the
superparticle mass spectrum, but  is generically bounded as
\begin{eqnarray}
    \label{eq:sigma_ann}
    \sigma_{\rm ann} \lesssim  \frac{\alpha_{1}}{m_{\chi}^{2}},
\end{eqnarray}
 with the coupling $\alpha_{1} \sim
10^{-2}$. Then, the ratio of the annihilation rate $\Gamma_{\rm ann} =
n_{\rm LSP}\sigma_{\rm ann}$, to the Hubble parameter $H \sim
T^{2}/\mG$ is bounded as 
\begin{eqnarray}
    \label{eq:gamma_to_H}
    \frac{\Gamma_{\rm ann}}{H} < \sigma_{\rm ann} \
    y_{3/2}^{\rm max} \ s \ \frac{\mG} {T^{2}} 
    = 10^{-4} \kakko{\frac{m_{\chi}} {100 \ \gev}}^{-3}
    \kakko{\frac{T}{\mev}}. 
\end{eqnarray}
We find $\Gamma_{\rm ann}/H < 1$ at the temperature of gravitino decay.


The LSPs may also be produced by the direct modulus decay into the
superparticles in the MSSM. This issue was investigated in
Ref.~\cite{Moroi:1999zb}. It turned out that the result depends on the
couplings of the modulus field to the MSSM fields. 
Consider, for instance, the case where the modulus field couples to
gauge multiplets through gauge kinetic function. Then the modulus decay into
a pair of gauginos receives chirality suppression, and the branching ratio 
is suppressed by $(m_{\tilde g}/m_{\phi})^2$, with $m_{\tilde g}$ being the
gaugino mass~\cite{Moroi:1999zb}. A rough estimate of the number density of
the LSPs produced by this decay gives
\begin{eqnarray}
   \frac{n_{\rm LSP}}{s} \sim {\rm Br}(\phi \rightarrow {\tilde g} {\tilde g})
                       \frac{T_R}{m_{\phi}}
                   \sim \left(\frac{m_{\tilde  g}}{m_{\phi}}\right)^2
                \left(\frac{m_{\phi}}{M_G}\right)^{1/2},
\end{eqnarray}
which is negligibly small for the range of the mass parameters of our concern.
Inclusion of the moduli coupling to quark/lepton chiral multiplets in
K\"{a}hler potential does not change the result, because the latter decay is
further suppressed by the fourth power of squark/slepton masses. 
In this paper, we assume that this is
the case, and we do not consider the direct production of the LSPs by
the modulus decay. From the above discussions about LSP, one can see
that our treatment gives conservative limits.

We now depict our constraints on $\mgrav$ and $\mphi$.  First in order
to clarify the influences on cosmology from the gravitino produced
only by the modulus decay, {\it i.e.}, from $y_{3/2,\phi}(t_{d})$ and
$y_{3/2,\rm th}(t_{d})$, for the moment we assume that $y_{3/2,\rm
inf}$ is negligibly small. This condition is represented by $J <
J_{\rm cr}$. If the reheating temperature after  inflation $T_{R,\rm
inf}$ is not so high, {\it i.e.}, $\lesssim 10^{12}$ GeV, this
situation is realized with $\phi_{i} \sim \mG$. Then $y_{3/2,\rm inf}$
is entirely diluted by the late-time entropy production caused by the
moduli decay.  Using the observational upper bounds in
(\ref{eq:y32_upper_photodis}) and (\ref{eq:yLSP_upper}), we can get
constraints on modulus mass $\mphi$ as a function of gravitino mass
$\mgrav$. From the expressions of $y_{3/2}(t_{d})$ in
Eq.~(\ref{direct})~and~Eq.~(\ref{thermal}), we can get the lower and
upper bounds on $\mphi$, respectively.

\begin{figure}[hp]
\begin{center}
    \epsfxsize=1.0\textwidth\epsfbox{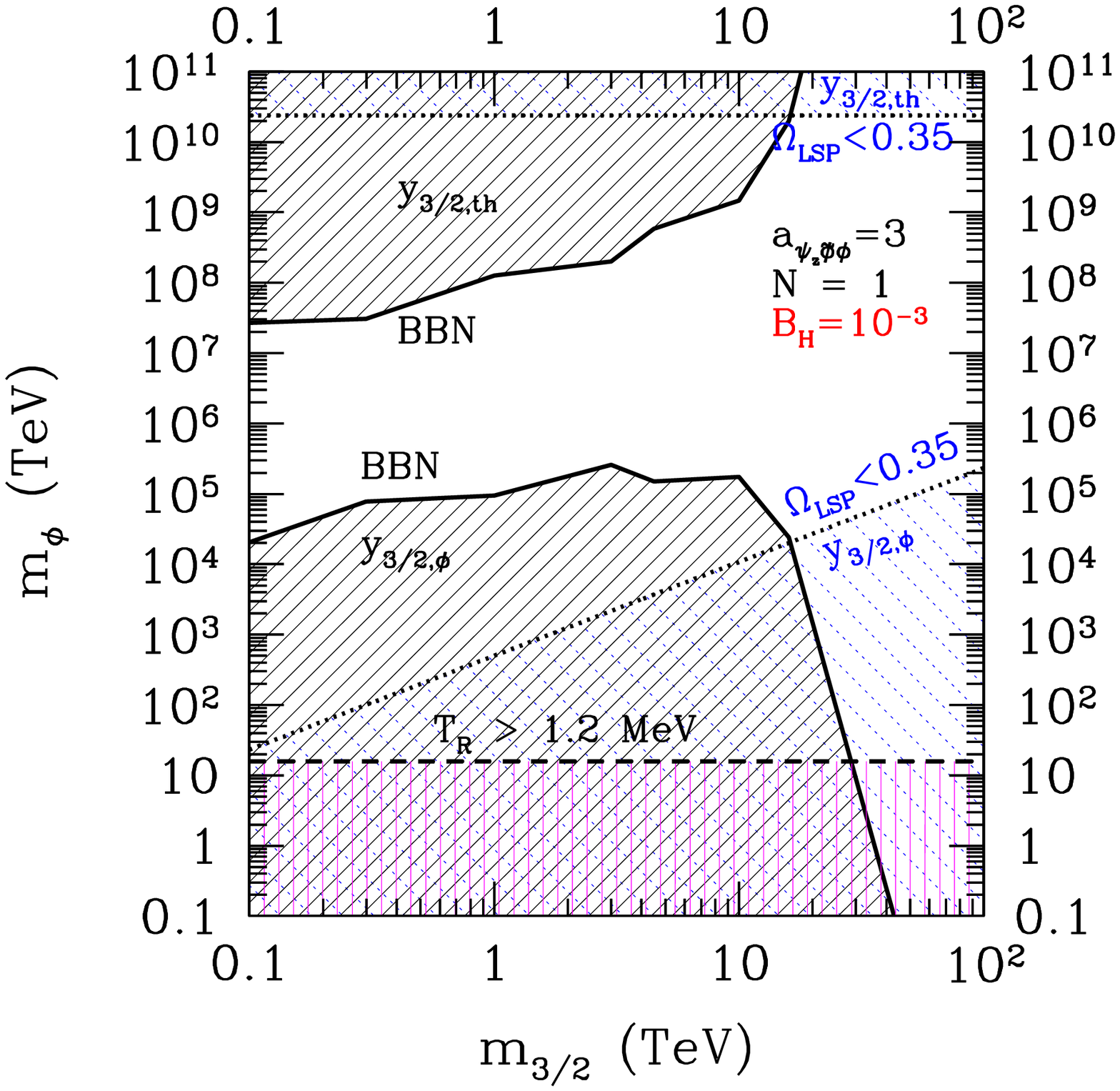}
\caption{
Constraints on modulus mass $\mphi$ as a function of gravitino mass
$\mgrav$ in case that the hadronic branching ratio is $B_{H} =
10^{-3}$. The solid lines denote the limits which come from the
photodissociation and hadro-dissociation  effects on light elements by
the decaying gravitino (BBN). The dotted lines represent constraints
on the energy density for LSPs, which are produced by decaying
gravitinos, to satisfy $\Omega_{\rm LSP} < 0.35$ for $m_{\rm LSP} =
100$ GeV. The dashed line denotes the lower bound on $\mphi$ arising
from the lower bound on the reheating temperature after modulus decay
to thermalize the neutrino background for successful BBN ($T_{R} >
1.2~\mev$). Here we adopted $\acoup = 3$ and $N = 1$. We also assumed
that the abundance of the gravitinos produced through the reheating
process after the primordial inflation are negligible. This
corresponds to the case that the reheating temperature after
primordial inflation is sufficiently low, $T_{R}^{\rm inf} = 0 -
10^{12}$ GeV, or $J < J_{\rm cr}$.
}
\label{fig:m32_mphi_photino}
\end{center}
\end{figure}

In Fig.~\ref{fig:m32_mphi_photino}, we plot the resultant constraints
on $\mphi$ as a function of $\mgrav$ for the case $y_{3/2,\rm inf}=0$
and the hadronic branching ratio $B_{H} = 10^{-3}$. Here we have taken
$\acoup = 3$  and $N = 1$ in Eq.~(\ref{eq:total}). We find that the
reheating requirement (\ref{eq:lower_bound}) gives a milder lower
bound on $\mphi$. From Fig.~\ref{fig:m32_mphi_photino}, we see that
the  modulus mass of the weak scale is excluded for gravitino mass of
$\mgrav$ = 0.1~--~100~$\tev$. In addition, it is interesting that we
can obtain the upper bound on $\mphi$  by this type of cosmological
arguments.  Finally we comment on the dependence of our constraints in
Fig.~\ref{fig:m32_mphi_photino} on $N$. Since $N$ appears only in the
form $N^{1/2} \mphi^{3/2}$ in all of the relevant expressions in
Eqs.~(\ref{eq:TR}),~(\ref{direct})~and~(\ref{thermal}), the
constraints for $N$ other than $N = 1$ can easily be read off by
replacing $\mphi$ in the vertical axis by $N^{1/3} \mphi$ in
Fig.~\ref{fig:m32_mphi_photino}.

\begin{figure}[hp]
\begin{center}
\epsfxsize=0.9\textwidth\epsfbox{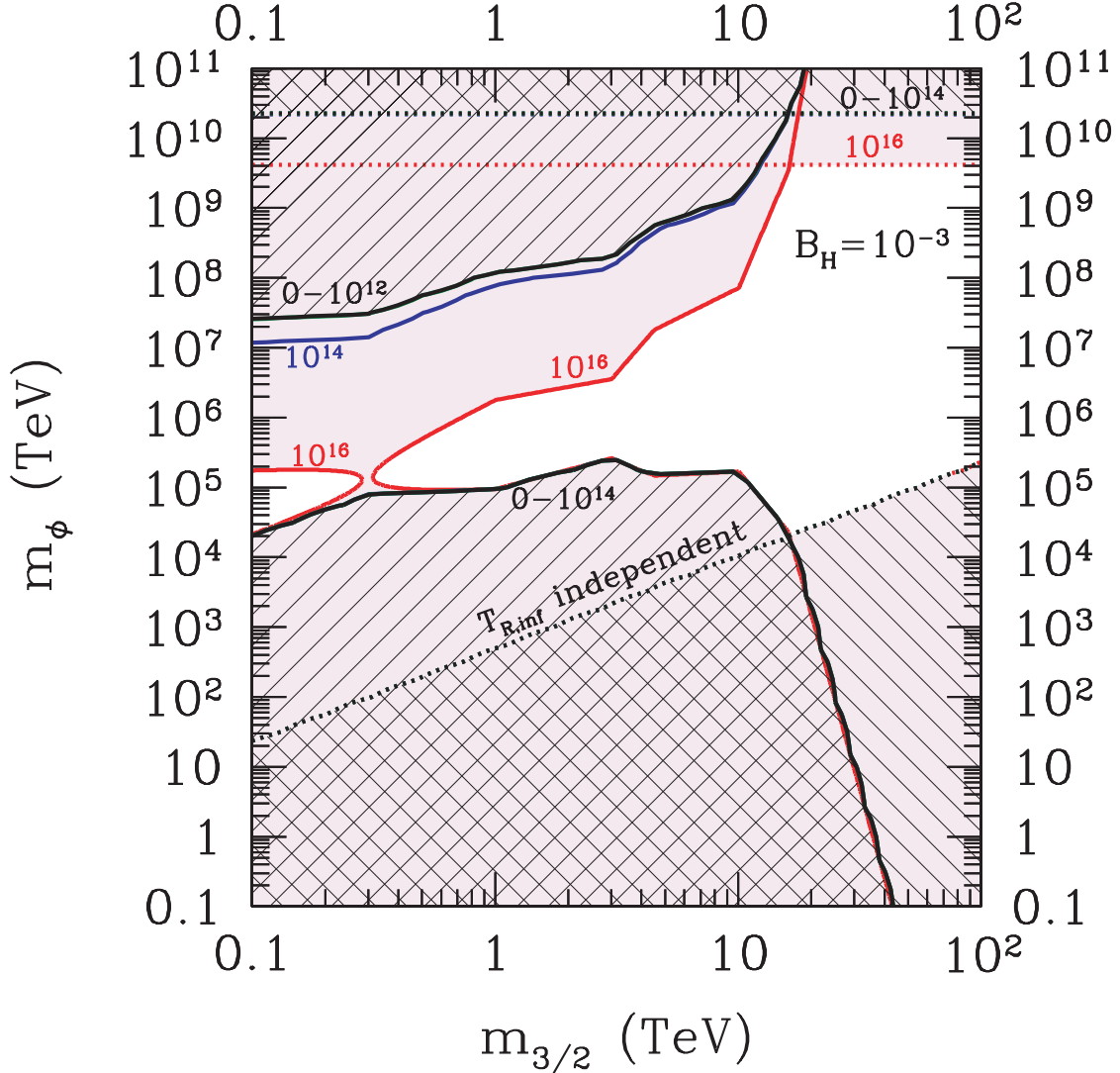}
\caption{
Same as Fig.~1, but for various reheating temperatures after
inflation, $T_{R,\rm inf}$ = 10$^{12}$, 10$^{14}$ and 10$^{16}$ GeV,
which are noted close to the corresponding lines.   The extent of
oblique lines coincides exactly with those excluded from the
constraints by BBN and LSP in Fig.~1, which correspond to the limiting
cases that $T_{R,\rm inf}$ is sufficiently low, $T_{R,\rm inf}= 0 -
10^{12}$ GeV.  The shadowed region corresponds to the excluded area
for the highest possible reheating temperature, $T_{R,\rm
inf}=10^{16}$ GeV. The shadowed region corresponds to the excluded
area for the highest possible reheating temperature, $T_{R,\rm
inf}=10^{16}$ GeV.  Here we have assumed that $\phi_i=\mG$ but other
cases can be easily read off from the fact that each line in the
figure is determined by the value of $J$.
}
\label{fig:prim_photino}
\end{center}
\end{figure}

\begin{figure}[hp]
\begin{center}
\epsfxsize=1.0\textwidth\epsfbox{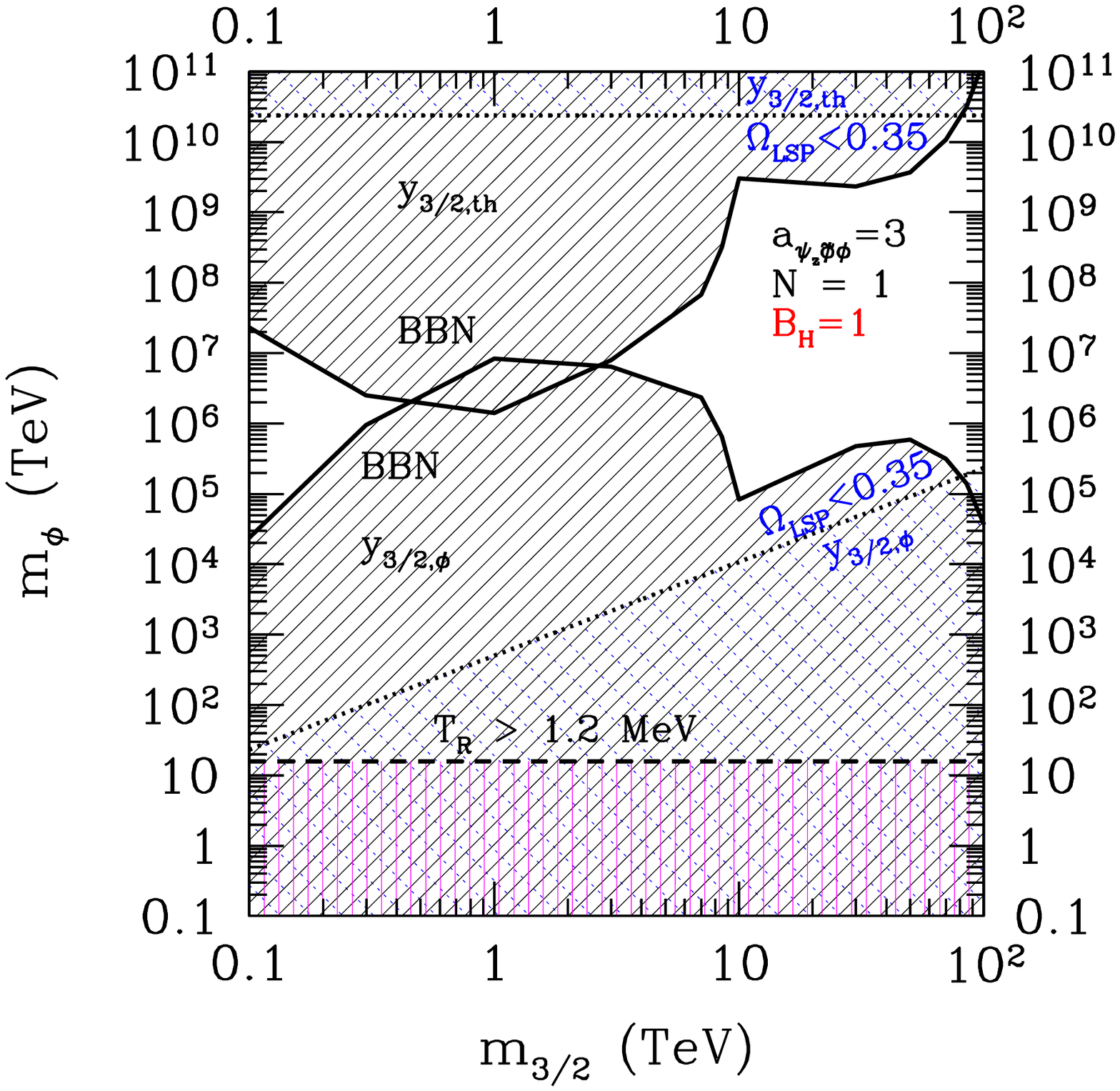}
\caption{
Same as Fig.~1, but for $B_H=1$
}
\label{fig:m32_mphi_gluino}
\end{center}
\end{figure}

\begin{figure}[hp]
\begin{center}
\epsfxsize=0.9\textwidth\epsfbox{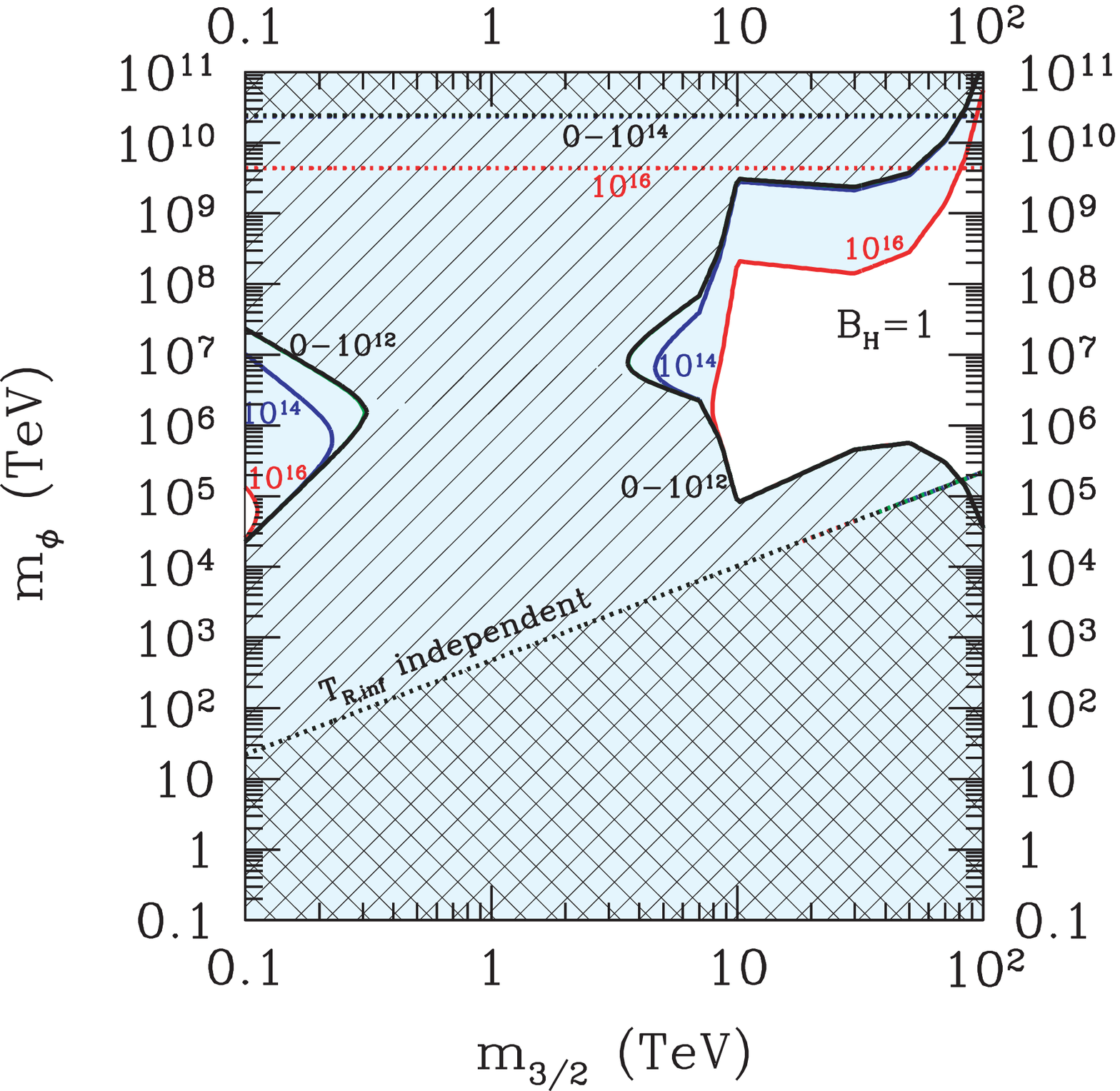}
\caption{
Same as Fig.~2, but for $B_H=1$
}
\label{fig:prim_gluino}
\end{center}
\end{figure}

Next we  discuss the  more general situation  where  $y_{3/2,\rm inf}$
can also be important with $J \gtrsim J_{\rm cr}$.  We  plot the
constraints on  $\mphi$ as a function of $\mgrav$ in
Fig.~\ref{fig:prim_photino} in the case $B_{H} = 10^{-3}$. The extent
of oblique lines coincides exactly with those excluded from the
constraints by BBN and LSP in Fig.~1, which correspond to the limiting
cases that $T_{R,\rm inf}$ is sufficiently  low such as $T_{R}^{\rm
inf} = 0 - 10^{12}$ GeV, namely, $J< J_{\rm cr}$.  For higher
reheating temperatures {\it e.g.}, $T_{R,\rm inf}$ = 10$^{14}$ --
10$^{16}$ GeV, larger parameter regions are additionally excluded.
The shadowed region corresponds to the excluded region for $T_{R,\rm
inf}=10^{16}$ GeV with $\phi_i=\mG$ or for $J=10^6$.  In a similar
fashion, the constraints for the case $B_{H}=1$ are depicted in
Fig.~\ref{fig:m32_mphi_gluino} and Fig.~\ref{fig:prim_gluino} for $J <
J_{\rm cr}$ and $J \gtrsim J_{\rm cr}$, respectively. From these
figures, we see that larger regions are excluded for $B_{H}=1$.

It is important to note that although the excluded region becomes
broader for higher reheating temperature, we still have a fairly large
{\it allowed} region in our parameter space even for the highest
possible reheating temperature $T_{R,\rm inf}=10^{16}$ GeV thanks to
the dilution of primordial gravitinos by the entropy production
associated with modulus decay.  Thus the previous upper bound on
$T_{R,\rm inf}$ in  (\ref{eq:y32_upper_tr_inf}) can easily be relaxed
if we consider the decaying moduli. In Table~1 (Table~2) we show the
allowed values of $\mphi$ for various $\mgrav$ and $T_{R,\rm inf}$ for
the case  $B_{H}=10^{-3}$ ($B_{H}=1$).

\section{Conclusion}
\label{sec:conclusion}

We have studied the effects of decaying modulus oscillation on the
cosmological gravitino problem.  We have considered a new direct
production mechanism of gravitinos from modulus decay, namely, a decay
mode of modulus into a gravitino and a modulino.  The width of this
decay mode can be larger than the other mode into two gravitinos which
has been studied in \cite{Hashimoto:1998mu}, if the coupling constant
is of the same order of magnitude with $\acoup \gtrsim 1$.

Comparing our yield of gravitinos with the constraints imposed by BBN
and the relic LSPs, which are decay products of gravitinos, we have
obtained a constraint on the masses of gravitinos and modulus.  As a
result we have found that due to the above-mentioned direct production
of gravitinos from decaying modulus, the modulus mass with $\mphi <
10^4$TeV is excluded, even when the branching ratio into hadrons is
minimal.

On the other hand, we have also found that wide range of $\mgrav$ and
$\mphi$ are still allowed even if the reheating temperature after
inflation is as high as $T_{R,\rm inf}=10^{16}$ GeV and the effects on
the hadronic decay of the gravitinos are taken into account, thanks to
the dilution of primordial gravitinos due to the entropy production
associated with modulus decay.

Thus in order to study cosmological consequences of gravitinos, it is
important to analyze not only their abundance right after inflation
but also their subsequent dilution due to late-time entropy
production, as well as late-time production from scalar condensates
with only gravitationally suppressed interactions including a dilaton
and a Polonyi field.

\section*{Acknowledgments}

This work was partially supported by the JSPS Grants-in Aid of the
Ministry of Education, Science, Sports, and Culture of Japan
No.~15-03605 (KK), No.~13640285 (JY), and No.12047201 (MY).

%

%
%

%
%

%
%



\begin{table}[htbp]
    \centering
    \begin{tabular}{cccccc}
        & $\mgrav$ & $T_{R, \rm inf}$= 0-$10^{12}$ GeV & $T_{R, \rm
        inf}= 10^{14}$ & $T_{R, \rm inf}= 10^{16}$ GeV & \\ 
        \hline
        & 0.1 TeV
        & $2 \times 10^4 - 3 \times 10^{7}$ TeV
        & $2 \times 10^4 -  1 \times 10^{7}$ TeV
        & $2 \times 10^4 -  2\times 10^{5}$ TeV
        & \\ 
        & 0.3 TeV
        & $8 \times 10^4 - 3 \times 10^{7}$ TeV
        & $8 \times 10^4 - 1 \times 10^{7}$ TeV
        & excluded
        & \\ 
        & 1 TeV
        & $1 \times 10^5 - 1 \times 10^{8}$ TeV
        & $1 \times 10^{5} - 8 \times 10^{7}$ TeV
        & $1 \times 10^5 - 2 \times 10^{6}$ TeV
        & \\ 
        & 3 TeV
        & $3 \times 10^5 - 2\times 10^{8}$ TeV
        & $3 \times 10^5 - 1\times 10^{8}$ TeV
        & $3 \times 10^5 - 3 \times 10^{6}$ TeV
        & \\ 
        &  10 TeV
        & $ 2 \times 10^5 - 2 \times 10^{9}$ TeV
        & $ 2 \times 10^5  - 2 \times 10^{9}$ TeV
        & $ 2 \times 10^5  - 7 \times 10^{7}$ TeV
        & \\ 
        &  30 TeV
        & $ 5 \times 10^4 - 2 \times 10^{10}$ TeV
        & $ 5 \times 10^4  - 2 \times 10^{10}$ TeV
        & $ 5 \times 10^4  - 4 \times 10^{9}$ TeV
        & \\ 
        &  100 TeV
        & $ 2 \times 10^5 - 2 \times 10^{10}$ TeV
        & $ 2 \times 10^5  - 2 \times 10^{10}$ TeV
        & $ 2 \times 10^5  - 4 \times 10^{9}$ TeV
        & \\ 
     \end{tabular}
    \caption{Allowed values of moduli mass $\mphi$ for various
        $\mgrav$ and  $T_{R, \rm inf}$ for $B_{H}=10^{-3}$.} 
    \label{tab:BHm3}
\end{table}

\begin{table}[htbp]
    \centering
    \begin{tabular}{cccccc}
        & $\mgrav$ & $T_{R, \rm inf}$= 0-$10^{12}$ GeV & $T_{R, \rm
        inf}= 10^{14}$ & $T_{R, \rm inf}= 10^{16}$ GeV & \\ 
        \hline
        & 0.1 TeV
        & $2 \times 10^4 - 2 \times 10^{7}$ TeV
        & $2 \times 10^4 -  1 \times 10^{7}$ TeV
        & $2 \times 10^4 -  1\times 10^{5}$ TeV
        & \\ 
        & 0.3 TeV
        & $1 \times 10^6 - 2 \times 10^{6}$ TeV
        & excluded
        & excluded
        & \\ 
        & 1 TeV
        & excluded
        & excluded
        & excluded
        & \\ 
        & 3 TeV
        & excluded
        & excluded
        & excluded
        & \\ 
        &  10 TeV
        & $ 1 \times 10^5 - 3 \times 10^{9}$ TeV
        & $ 1 \times 10^5  - 3 \times 10^{9}$ TeV
        & $ 1 \times 10^5  - 2 \times 10^{8}$ TeV
        & \\ 
        &  30 TeV
        & $ 5 \times 10^5 - 2 \times 10^{9}$ TeV
        & $ 5 \times 10^5  - 2 \times 10^{9}$ TeV
        & $ 5 \times 10^5  - 4 \times 10^{9}$ TeV
        & \\ 
        &  100 TeV
        & $ 2 \times 10^5 - 2 \times 10^{10}$ TeV
        & $ 2 \times 10^5  - 2 \times 10^{10}$ TeV
        & $ 2 \times 10^5  - 4 \times 10^{9}$ TeV
        & \\ 
     \end{tabular}
    \caption{Same as Table~1, but for $B_{H}=1$.} 
    \label{tab:BH1}
\end{table}


\begin{references}

\bibitem{Weinberg:zq}
S.~Weinberg,
Phys.\ Rev.\ Lett.\  {\bf 48}, 1303 (1982).

\bibitem{Krauss:1983ik}
L.~M.~Krauss,
Nucl.\ Phys.\ B {\bf 227}, 556 (1983).


\bibitem{Lindley:1984bg}
D.~Lindley,
Astrophys.\ J.\  {\bf 294} (1985) 1.


\bibitem{Khlopov:pf}
M.~Y.~Khlopov and A.~D.~Linde,
Phys.\ Lett.\ B {\bf 138}, 265 (1984);
F.Balestra, G.Piragino, D.B.Pontecorvo, M.G.Sapozhnikov,
I.V.Falomkin, M.Yu.Khlopov,
Sov. J. Nucl. Phys. 39 (1984)  626;
M.Yu. Khlopov, Yu.L.Levitan, E.V.Sedelnikov and I.M.Sobol,
Phys. Atom.Nucl.  57 (1994) 1393;
M.~Y.~Khlopov,
``Cosmoparticle Physics,''
(Singapore: World Scientific, 1999)


\bibitem{Ellis:1984eq} J.~R.~Ellis, J.~E.~Kim and D.~V.~Nanopoulos,
Phys.\ Lett.\ B {\bf 145}, 181 (1984).


\bibitem{Juszkiewicz:gg}
R.~Juszkiewicz, J.~Silk and A.~Stebbins,
Phys.\ Lett.\ B {\bf 158}, 463 (1985).


\bibitem{Ellis:1984er}
J.~R.~Ellis, D.~V.~Nanopoulos and S.~Sarkar,
Nucl.\ Phys.\ B {\bf 259} (1985) 175.

\bibitem{Audouze:be}
J.~Audouze, D.~Lindley and J.~Silk,
Astrophys.\ J.\  {\bf 293}, L53 (1985)
;
D.~Lindley,
Phys.\ Lett.\ B {\bf 171} (1986) 235.



\bibitem{Kawasaki:1986my}
M.~Kawasaki and K.~Sato,
Phys.\ Lett.\ B {\bf 189}, 23 (1987).


\bibitem{Scherrer:1987rr}
R.~J.~Scherrer and M.~S.~Turner,
Astrophys.\ J.\  {\bf 331} (1988) 19.


\bibitem{Dominguez:1987}
R. Dominguez-Tenreiro, 
Astrophys.\ J.\  {\bf 313},  523 (1987).


\bibitem{Reno:1987qw}
M.~H.~Reno and D.~Seckel,
Phys.\ Rev.\ D {\bf 37} (1988) 3441.

\bibitem{Dimopoulos:1987fz}
S.~Dimopoulos, R.~Esmailzadeh, L.~J.~Hall and G.~D.~Starkman,
Astrophys.\ J.\  {\bf 330}, 545 (1988)
;
Phys. Rev. Lett. 60, (1988) 7
;
Nucl.\ Phys.\ B {\bf 311} (1989) 699.

\bibitem{Ellis:1990nb}
J.~R.~Ellis, G.~B.~Gelmini, J.~L.~Lopez, D.~V.~Nanopoulos and S.~Sarkar,
Nucl.\ Phys.\ B {\bf 373}, 399 (1992).



\bibitem{Kawasaki:1994af}
M.~Kawasaki and T.~Moroi,
Prog.\ Theor.\ Phys.\  {\bf 93} (1995) 879
[arXiv:hep-ph/9403364]
;
Astrophys.\ J.\  {\bf 452}, 506 (1995)
[arXiv:astro-ph/9412055].


\bibitem{Kawasaki:1994bs}
M.~Kawasaki and T.~Moroi,
Phys.\ Lett.\ B {\bf 346}, 27 (1995)
[arXiv:hep-ph/9408321].

\bibitem{Protheroe:dt}
R.~J.~Protheroe, T.~Stanev and V.~S.~Berezinsky,
Phys.\ Rev.\ D {\bf 51}, 4134 (1995)
[arXiv:astro-ph/9409004].


\bibitem{Holtmann:1998gd}
E.~Holtmann, M.~Kawasaki, K.~Kohri and T.~Moroi,
Phys.\ Rev.\ D {\bf 60}, 023506 (1999)
[arXiv:hep-ph/9805405].


\bibitem{Jedamzik:1999di}
K.~Jedamzik,
Phys.\ Rev.\ Lett.\  {\bf 84}, 3248 (2000)
[arXiv:astro-ph/9909445].


\bibitem{Kawasaki:2000qr}
M.~Kawasaki, K.~Kohri and T.~Moroi,
Phys.\ Rev.\ D {\bf 63}, 103502 (2001)
[arXiv:hep-ph/0012279].


\bibitem{Kohri:2001jx}
K.~Kohri,
Phys.\ Rev.\ D {\bf 64} (2001) 043515
[arXiv:astro-ph/0103411].


\bibitem{Cyburt:2002uv}
R.~H.~Cyburt, J.~R.~Ellis, B.~D.~Fields and K.~A.~Olive,
Phys.\ Rev.\ D {\bf 67}, 103521 (2003)
[arXiv:astro-ph/0211258].


\bibitem{Kawasaki:2004yh}
M.~Kawasaki, K.~Kohri and T.~Moroi,
arXiv:astro-ph/0402490.




\bibitem{Pagels:ke}
H.~Pagels and J.~R.~Primack,
Phys.\ Rev.\ Lett.\  {\bf 48}, 223 (1982).


\bibitem{Berezinsky:kf}
V.~S.~Berezinsky,
Phys.\ Lett.\ B {\bf 261}, 71 (1991).


\bibitem{Moroi:1993mb}
T.~Moroi, H.~Murayama and M.~Yamaguchi,
Phys.\ Lett.\ B {\bf 303}, 289 (1993).


\bibitem{Moroi:1995fs}
T.~Moroi,
arXiv:hep-ph/9503210.



\bibitem{Bolz:2000fu}
M.~Bolz, A.~Brandenburg and W.~Buchmuller,
Nucl.\ Phys.\ B {\bf 606}, 518 (2001)
[arXiv:hep-ph/0012052].


\bibitem{hybrid}
A.~D.~Linde,
\PLB{259}{38}{91};
Phys.\ Rev.\ D {\bf 49}, 748 (1994).

\bibitem{YY}
M. Kawasaki, M. Yamaguchi, and J. Yokoyama,
\PRDD{68}{023508}{03}; M.\ Yamaguchi and J.\ Yokoyama,
\PRDD{68}{123530}{03}; M.\ Yamaguchi and J.\ Yokoyama, hep-ph/0402282.



\bibitem{Coughlan:1983ci}
G.~D.~Coughlan, W.~Fischler, E.~W.~Kolb, S.~Raby and G.~G.~Ross,
Phys.\ Lett.\ B {\bf 131}, 59 (1983).

\bibitem{Banks:1993en}
T.~Banks, D.~B.~Kaplan and A.~E.~Nelson,
Phys.\ Rev.\ D {\bf 49}, 779 (1994)
[arXiv:hep-ph/9308292].

\bibitem{deCarlos:1993jw}
B.~de Carlos, J.~A.~Casas, F.~Quevedo and E.~Roulet,
Phys.\ Lett.\ B {\bf 318}, 447 (1993)
[arXiv:hep-ph/9308325].

\bibitem{MYY}
T.~Moroi, M.~Yamaguchi and T.~Yanagida,
Phys.\ Lett.\ B {\bf 342}, 105 (1995)
[arXiv:hep-ph/9409367].



\bibitem{Kawasaki:1995cy}
M.~Kawasaki, T.~Moroi and T.~Yanagida,
Phys.\ Lett.\ B {\bf 370}, 52 (1996)
[arXiv:hep-ph/9509399].

\bibitem{Moroi:1999zb}
T.~Moroi and L.~Randall,
Nucl.\ Phys.\ B {\bf 570}, 455 (2000)
[arXiv:hep-ph/9906527].


\bibitem{Kawasaki:1999na}
M.~Kawasaki, K.~Kohri and N.~Sugiyama,
Phys.\ Rev.\ Lett.\  {\bf 82}, 4168 (1999)
[arXiv:astro-ph/9811437];
Phys.\ Rev.\ D {\bf 62}, 023506 (2000)
[arXiv:astro-ph/0002127].






\bibitem{Hashimoto:1998mu}
M.~Hashimoto, K.~I.~Izawa, M.~Yamaguchi and T.~Yanagida,
Prog.\ Theor.\ Phys.\  {\bf 100}, 395 (1998)
[arXiv:hep-ph/9804411].

\bibitem{Giddings}
S.~B.~Giddings, S.~Kachru and J.~Polchinski,
Phys.\ Rev.\ D {\bf 66}, 106006 (2002)
[arXiv:hep-th/0105097].


\bibitem{Kaplunovsky:1993rd}
V.~S.~Kaplunovsky and J.~Louis,
Phys.\ Lett.\ B {\bf 306}, 269 (1993)
[arXiv:hep-th/9303040].




\bibitem{LythStewart}
D.~H.~Lyth and E.~D.~Stewart,
Phys.\ Rev.\ D {\bf 53}, 1784 (1996)
[arXiv:hep-ph/9510204].


\bibitem{Spergel:2003cb}
D.~N.~Spergel {\it et al.},
Astrophys.\ J.\ Suppl.\  {\bf 148}, 175 (2003)
[arXiv:astro-ph/0302209].

\bibitem{freedman:2001}
W. Freedman et al.,
Astrophys.\ J.\ {\bf 553}, 47 (2001).

\bibitem{Mather:1999}
J.C. Mather et al.,
Astrophys.\ J.\ {\bf 512}, 511 (1999).



\end{references}
\end{document}